\documentclass[twocolumn,aps,prl,showpacs,floats,letterpaper,floatfix,groupedaddress]{revtex4}

\usepackage{amssymb,amsmath}
\usepackage[dvips]{graphicx}

\usepackage{dcolumn,epsfig}

\newcommand{\beq}{\begin{equation}}
\newcommand{\eeq}{\end{equation}}
\newcommand{\bes}{\begin{subequations}}
\newcommand{\ees}{\end{subequations}}
\newcommand{\bea}{\begin{eqnarray}}
\newcommand{\eea}{\end{eqnarray}}
\newcommand{\ba}{\begin{array}}
\newcommand{\ea}{\end{array}}
\newcommand{\beqn}{\begin{eqnarray*}}
\newcommand{\eeqn}{\end{eqnarray*}}

\newcommand{\f}[2]{\frac{#1}{#2}}

\newcommand{\la}{\langle}
\newcommand{\dg}{\dagger}
\newcommand{\ra}{\rangle}

\def\nn{\nonumber}

\newlength{\sizeonefig}
\newlength{\sizetwofig}
\begin{document}

\title{Few-photon optical diode}

\author{Dibyendu Roy} 
\email{droy@physics.ucsd.edu}
\affiliation{Department of Physics, University of California-San Diego, La Jolla, California 92093-0319}

\begin{abstract}
We propose a novel scheme of realizing an optical diode at the few-photon level. The system consists of a one-dimensional waveguide coupled asymmetrically to a two-level system. The two or multi-photon transport in this system is strongly correlated. We derive exactly the single and  two-photon current and show that the two-photon current is asymmetric for the asymmetric  coupling. Thus the system serves as an optical diode which allows transmission of photons in one direction much more efficiently than the opposite. 
\end{abstract}

\vspace{0.5cm}

\pacs{: 03.65.Nk, 42.50.-p, 32.80.-t, 73.40.Ei}
\maketitle
A diode is an essential circuit element that allows unidirectional propagation of signal. Faraday rotator based on magneto-optic effect serves as an optical diode or optical isolator in most modern optical labs. There are also several proposed or demonstrated optical diodes relying on cascaded nonlinearities \cite{Opdiode}. In this paper we propose a novel scheme to realize an optical diode at the few-photon level by making photons strongly interacting. Different schemes \cite{Imamoglu97, Harris98, Kojima03, Koshino04, Birnbaum05} have been explored to produce strong nonlinearities at the few-photon level. Recently a new approach has been suggested by Shen and Fan (SF) \cite{Shen07} to create photon-photon nonlinear interactions by coupling a bare two-level system (TLS) with a one-dimensional continuum for photons, such as a photonic crystal waveguide. Local interactions at the TLS induce strong correlations between photons. Based on a similar mechanism of realizing nonlinear interactions, an all optical single-photon transistor has been proposed by using surface plasmon modes of a conducting nano-wire \cite{Chang07}.

\begin{figure}
\includegraphics[width=6.5cm]{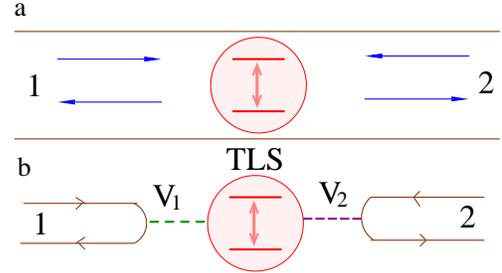}
\caption{(color online) (a) A schematic of the two-level system being directly coupled to photons in a one-dimensional wave-guide. (b) Unfolded model for the photons in the left (1) and right (2) side of the two-level system.}
\label{pl1}
\end{figure}

SF consider a system of a one-dimensional waveguide with left and right-moving photons side-coupled to a TLS. The two-photon transport  has been studied by them using a generalized Bethe-ansatz technique which properly takes into account the open boundary conditions of the problem \cite{Shen07}. It has been shown that the exact two-photon scattering eigenstates of this system include a two-photon bound state that passes through the TLS as a composite single particle. In this  paper we study a similar electro-optical system that consists of a TLS directly coupled to a 1D photonic waveguide (see top of Fig.\ref{pl1}). Also the coupling of the TLS with the waveguide is asymmetric. In a line-defect photonic waveguide the asymmetry in coupling may be realized by modulating the periodicity of photonic crystal at any side of the TLS locally. It is also possible to fabricate asymmetric coupling by working with a system of superconducting transmission line resonators and a dc-SQUID based charge qubit \cite{Wallraff04, Zhou08}, or a system consisting of a bare two-level emitter interacting with the surface-plasmon modes of a conducting nanowire \cite{Chang07, Akimov07, Chang2}. We construct the exact single and two-photon scattering states for our model by directly solving the single and two-particle Schr{\"o}dinger equations. Then we evaluate the single and two-particle current in the system. We find an interesting asymmetry in the two-photon current for asymmetric coupling between the TLS and the waveguide. This occurs due to a redistribution of energy and momentum of photons after scattering from the TLS in the absence of spatial symmetry. Thus, our model acts as an optical diode which permits photons to transmit in a direction more easily compared to the opposite direction. It is easier to achieve asymmetry in coupling for the direct-coupled TLS-waveguide system compared to the side-coupled TLS-waveguide.

\begin{figure}
\includegraphics[width=8.5cm]{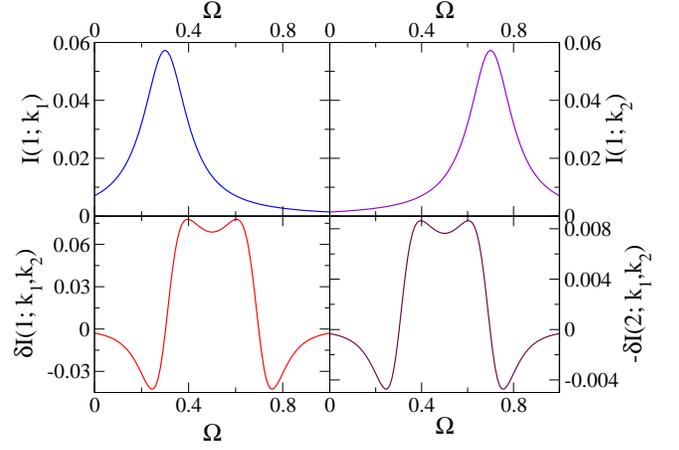}
\caption{(color online) Plot of the single photon current $[I(1;k_1),~I(1;k_2)]$, the two-photon current change $[\delta I(1;k_1,k_2),~\delta I(2;k_1,k_2)]$  with the transition energy $\Omega$. Here $k_1=0.3$, $k_2=0.7$, $V_1=0.45$, and $V_2=0.15$. 
}
\label{pl2}
\end{figure}

We consider a simple spin-boson type Hamiltonian  to describe the direct-coupled TLS-waveguide system. The Hamiltonian, $\mathcal{H}=\hbar \Omega \sigma_z/2+ \sum_{i,k} \hbar \omega_{ik} a^{\dg}_{ik}a_{ik}+\sum_{i,k}V_{ik}(a^{\dg}_{ik}+a_{ik})(\sigma_{+}+\sigma_{-})$, where the first term represents the free TLS with transition energy $\Omega$ (setting $\hbar=1$); the second term corresponds to the free photons with energy $\omega_{ik}$ in the left $(i=1)$ and the right $(i=2)$ side of the TLS, and the last term describes the interaction between the TLS and the photons. Here $a_{ik}~(a^{\dg}_{ik})$ is the annihilation (creation) operator for photons  in the $i$th side of the TLS and $V_{1k}~(V_{2k})$ is the coupling between the photon in the left (right) side of the TLS and the TLS. $\sigma_{-}=a_g^{\dagger}a_e~(\sigma_{+}=a_e^{\dagger}a_g)$ is the lowering (raising) ladder operator where $a_g^{\dagger} (a_e^{\dagger})$ is the creation operator of the ground (excited) state of the TLS. Then we unfold the model by performing standard manipulations for impurity models with $\Omega$ being away from the cut-off of the linear dispersion (see bottom of Fig.\ref{pl1}). Thus obtained the real-space Hamiltonian of the system as chiral 1D field theories is given by
\bea
\mathcal{H}&=&-i \sum_{\alpha=1,2}v^{\alpha}_g\int dx ~b^{\dagger}_{\alpha}(x)\partial_xb_{\alpha}(x)+\Omega a_e^{\dagger}a_e \nn \\
&+&(\tilde{V}_1b_1^{\dg}(0)\sigma_{-}+\tilde{V}_2b_2^{\dg}(0)\sigma_{-}+H.c.)~,
\label{Ham0}
\eea
where $v^{1}_g~[v^{2}_g]$ is the group velocity of the photons and $b^{\dg}_{1}(x)~[b^{\dg}_{2}(x)]$ is a photon creation operator at $x$ in the left (right) side of the TLS.  $V_{1k}~(V_{2k})$ is replaced by a frequency independent $\tilde{V}_1~ (\tilde{V}_2)$. 
Here $\sigma_{+}|n,-\ra=|n,+\ra$ and $\sigma_{+}|n,+\ra=0$, where $|n,\pm\ra$ represents the full system with $n$ photons and the TLS in the excited $(+)$ or ground $(-)$ state. Next we redefine the field operators in the two sides of the TLS, $c_{\alpha}(x)=\sqrt{v^{\alpha}_g}~b_{\alpha}(x)$. We assume here that $v^{\alpha}_g$ is real and $x-$independent. Thus we find
\bea
\mathcal{H}&=&-i\sum_{\alpha=1,2}\int dx ~c^{\dagger}_{\alpha}(x)\partial_xc_{\alpha}(x)+\Omega a_e^{\dagger}a_e \nn \\
&+&(V_1c_1^{\dg}(0)\sigma_{-}+V_2c_2^{\dg}(0)\sigma_{-}+H.c.)~,
\label{Ham}
\eea
with $V_{\alpha}=\tilde{V}_{\alpha}/\sqrt{v^{\alpha}_g}$. It shows that the renormalized coupling can be asymmetric for different group velocities of photons at the two sides of the TLS even for $\tilde{V}_1= \tilde{V}_2$. We define the current operator in the system as $I=-i[\mathcal{H}, N_1-N_2]/2$ where $N_1~(N_2)$ is the total number of photons in the left (right) side of the TLS. Then we find $I=i(V_1c^{\dg}_1(0)\sigma_{-}-V_2c^{\dg}_2(0)\sigma_{-}-H.c.)/2$. We want to find the expectation value of the current operator in the steady state. Thus we need to find the exact scattering eigenstates of the $\mathcal{H}$. We here observe that under the transformation, $c_{1}(x)=(V_1c_{e}(x)+V_2c_{o}(x))/V$ and $c_{2}(x)=(V_2c_{e}(x)-V_1c_{o}(x))/V$ with $V=\sqrt{V_1^2+V_2^2}$,
the Hamiltonian in Eq.(\ref{Ham}) breaks into two decoupled parts of even and odd field operators; i.e., $\mathcal{H}=\mathcal{H}_e+\mathcal{H}_o$ where 
$\mathcal{H}_e =-i \int dx~ c^\dg_{e}(x)\partial_xc_{e}(x) + \Omega a_e^{\dagger}a_e + V \big(c^\dg_{e}(0)\sigma_{-}+\sigma_{+}c_{e}(0)\big)~{\rm and}~\mathcal{H}_o =-i \int dx~ c^\dg_{o}(x)\partial_ xc_{o}(x)$. 
In the transformed basis, the $\mathcal{H}_e$  contains the interaction term between the even modes with the TLS while the $\mathcal{H}_o$ is just the kinetic energy of the noninteracting odd modes. The current operator in the even-odd basis is given by
\bea
I= \f{i}{2}\Big(\f{V_1^2-V_2^2}{V}c^{\dg}_e(0)\sigma_{-}+\f{2V_1V_2}{V}c^{\dg}_o(0)\sigma_{-}-H.c.\Big)~.\label{curreo}
\eea
{\it Single-photon dynamics:} First we construct the exact single-photon scattering states of the full system and  derive the corresponding single-photon current for a photon incoming from the left or the right side of the TLS. The single-photon eigenstates of the full system are 
\bea
\int \f{dx}{\sqrt{2\pi}} \{A_1[g_k(x)c^{\dg}_e(x)+\delta(x)e_k\sigma_{+}]+B_1h_k(x)c^{\dg}_{o}(x)\}|0,-\ra,\label{sinstate}
\eea
with $A_1$ and $B_1$ being arbitrary constants to satisfy the nonequilibrium boundary conditions, i.e., the incident photon is coming from which side of the TLS. $e_k$ is the excitation amplitude of the TLS. We find different coefficients in the eigenstates by solving the following equations of motion obtained from the stationary single photon Schr{\"o}dinger equation.
\bea
-i\partial_xg_{k}(x)-k g_{k}(x)+Ve_k\delta(x)&=&0\nn \\
(\Omega-k)e_k+Vg_k(x)\delta(x)&=&0\nn \\
-i\partial_xh_{k}(x)-k h_{k}(x)&=&0
\eea
where $g_k(0)=[g_k(0-)+g_k(0+)]/2$ and $g_k(x)$ is a single photon plane wave for $x<0$. We find $g_k(x)=e^{ikx}\Big[\theta(-x)+r_k\theta(x)\Big],~h_k(x)=e^{ikx},~e_k=V/(k-\Omega+iV^2/2),$ and $r_k=e_k/e_k^*$, where $\theta(x)$ is the step function. We define the single-photon scattering states as $|1;k\ra~(|2;k\ra)$ for those with an incoming photon of momentum $k$ from the left (right) side of the TLS. By choosing $A_1=V_1/V,~B_1=V_2/V~(A_1=V_2/V,~B_1=-V_1/V)$ in Eq.(\ref{sinstate}) we find the state $|1;k\ra~(|2;k\ra)$. We then evaluate the steady state single photon current in $|1;k\ra$.
\bea
I(1;k)=\la 1;k|I|1;k\ra&=&-\f{1}{\pi}\f{V_1^2V_2^2}{V^3}{\rm Im}[e_{k_1}]\nn \\
&=&\f{2}{\pi}\f{\Gamma_1\Gamma_2}{(k-\Omega)^2+(\Gamma_1+\Gamma_2)^2}\label{sincurr}
\eea
with $\Gamma_{\alpha}=V_{\alpha}^2/2$. 
As expected in the absence of any background phase shifts (which may arise if the TLS is embedded in a cavity) the single-photon current shows a Breit-Wigner-like (i.e., Lorentzian) line shape around the resonance $k=\Omega$ (see top two plots in Fig.\ref{pl2}). For symmetric coupling, i.e., $\Gamma_1=\Gamma_2=\Gamma$, a single photon at resonance frequency is completely transmitted from one side of the TLS to the other side. Note here that the transmission and reflection coefficient for the direct-coupled model are similar to the reflection and transmission coefficient of the side-coupled model \cite{Shen07}, a fact which also occurs for waveguide-resonator systems \cite{Xu00}. We also evaluate $\la 2;k|I|2;k\ra$ and we find $|\la 1;k|I|1;k\ra|=|\la 2;k|I|2;k\ra|$ for arbitrary $V_1,V_2$, i.e, the current is the same in magnitude for an incident photon coming from either side of the TLS. This even can be understood from the current expression in Eq.(\ref{sincurr}) which remains the same under exchange of $V_1$ and $V_2$.   

{\it Two-photon dynamics and optical diode:} Next we construct the two-photon scattering eigenstates for this model. This is a nontrivial task. 
The two-photon incoming state for both the photons being incident from the left side of the TLS is given by
\bea
\int dx_1 dx_2 \f{1}{2\pi \sqrt{2}}\phi_{\bf k}(x_1,x_2)\f{1}{\sqrt{2}}c^{\dg}_1(x_1)c^{\dg}_1(x_2)|0,-\ra~, \label{instate}
\eea
where $\phi_{\bf k}(x_1,x_2)=(e^{ik_1x_1+ik_2x_2}+e^{ik_1x_2+ik_2x_1})$ with ${\bf k}=(k_1,k_2)$. We decompose the two-photon incoming state into $ee,~oo$, and $eo$ subspaces and determine the full scattering eigenstates in the different subspaces separately. Note that to calculate the expectation value of the current operator in Eq.(\ref{curreo}), we need to express the two-photon scattering eigenstate in the space of free photons as well as the TLS \cite{Nishino09}. This is similar to the single photon scattering states in (\ref{sinstate}). Thus we differ here from the approach of SF \cite{Shen07} where the outgoing scattering states are expressed by a complete orthonormal basis spanning the free two-photon Hilbert space. 

The two or multi-photon transport in one-dimensional waveguides for the direct-coupled TLS model is strongly correlated. The local interaction at the TLS induces strong correlations between photons by preventing multiple occupancies of photons at the TLS. Thus the Hamiltonian in Eq.\ref{Ham0} is equivalent to a bosonic version of the popular single-impurity Anderson model with an infinite repulsive interaction between photons at the TLS. When photons are incident upon the TLS, one of the photon gets absorbed and re-emitted by the TLS and then that interferes with the other photons in the waveguide. By the process of absorption and spontaneous emission the TLS mediates strong interactions between photons.
The general two-photon scattering states are of the form:
\bea
&&\int\Big[A_2\big\{g(x_1,x_2)\f{1}{\sqrt{2}}c^{\dg}_e(x_1)c^{\dg}_e(x_2)+e(x_1)\delta(x_2)c^{\dg}_e(x_1)\sigma_{+}\big\}\nn \\&&+B_2\big\{j(x_1;x_2)c^{\dg}_e(x_1)c^{\dg}_o(x_2)+f(x_1)\delta(x_2)c^{\dg}_o(x_1)\sigma_{+}\big\}\nn \\ &&+C_2~h(x_1,x_2)\f{1}{\sqrt{2}}c^{\dg}_o(x_1)c^{\dg}_o(x_2)\Big] dx_1dx_2|0,-\ra
\label{wavefn}
\eea 
where $g(x_1,x_2)=g(x_2,x_1)$ and $h(x_1,x_2)=h(x_2,x_1)$ due to Bose statistics of photons. $e(x)~(f(x))$ is the probability amplitude of one photon in the $e~(o)$-subspace while the TLS in the excited state. Here $A_2,B_2$ and $C_2$ identify the boundary conditions for the incoming photons. Like the single photon case we here determine all the functions in Eq.(\ref{wavefn}) using the two-photon Schr{\"o}dinger equation with the total energy of the two photons $E=k_1+k_2$. Thus, we obtain the following linear equations.
\bea
\Big(-i\partial_{x_1}-i\partial_{x_2}-E\Big)g(x_1,x_2)~~~~~~~~~~~~~~~~~ \nn \\+\f{V}{\sqrt{2}}[e(x_1)\delta(x_2)+\delta(x_1)e(x_2)]&=&0 \nn \\
\Big(-i\partial_x-E+\Omega\Big)e(x)+\f{V}{\sqrt{2}}[g(0,x)+g(x,0)]&=&0 \nn \\
\Big(-i\partial_{x_1}-i\partial_{x_2}-E\Big)j(x_1;x_2)+V\delta(x_1)f(x_2)&=&0\nn\\
\Big(-i\partial_x-E+\Omega\Big)f(x)+Vj(0;x)&=&0 \nn \\
\Big(-i\partial_{x_1}-i\partial_{x_2}-E\Big)h(x_1,x_2)&=&0\nn \label{Se}
\eea
We solve the above equations with the initial conditions that the functions $g(x_1,x_2), ~j(x_1;x_2)$, and $h(x_1,x_2)$ in the region $x_1,x_2<0$ satisfy the form of the corresponding plane wave-functions given in Eq.(\ref{instate}). We also use, $g(0,x)=g(x,0)\equiv [g(0+,x)+g(0-,x)]/2$, $j(0,x)\equiv[j(0+,x)+j(0-,x)]/2$.  
In our solution we explicitly separate out the two-photon bound state which arises in the $ee$-subspace. We find
\bea
g(x_1,x_2)&=&\f{1}{2\pi\sqrt{2}}g_{k_1}(x_1)g_{k_2}(x_2)+\f{V^2}{\sqrt{2}\pi}e_{k_1}e_{k_2}e^{i(\Omega-iV^2/2)x_2}\nn \\&&e^{i(E-\Omega+iV^2/2)x_1}\theta(x_1-x_2)\theta(x_2)+(1 \leftrightarrow 2)\nn\\
e(x)&=&\f{1}{2\pi}\Big( g_{k_1}(x)e_{k_2}+g_{k_2}(x)e_{k_1}\Big)\nn \\&& +\f{iV}{\pi}e_{k_1}e_{k_2}e^{i(E-\Omega+iV^2/2)x}\theta(x) \nn \\
j(x_1;x_2)&=&\f{1}{2\pi}\Big( g_{k_1}(x_1)h_{k_2}(x_2)+g_{k_2}(x_1)h_{k_1}(x_2)\Big)\nn \\
f(x)&=&\f{1}{2\pi}\Big( e_{k_1}h_{k_2}(x)+e_{k_2}h_{k_1}(x)\Big)\nn \\
h(x_1,x_2)&=&\f{1}{2\pi\sqrt{2}}~\phi_{\bf k}(x_1,x_2)
\label{ampl}
\eea

The second terms in $g(x_1,x_2)$ and $e(x)$ are contributions from the two-photon bound state. Both of these terms decay to zero with $|x_1-x_2|$ or $|x|$ tending to infinity; this is an essential characteristic of the bound state. 
We emphasize that due to interactions at a localized region two photons can exchange energy and momentum between themselves with the constraint of a fixed total energy and momentum. This redistribution in energy creates a strong interaction between photons. 
 
Now we calculate the expectation value of the current operator in the two-photon scattering state $|1;k_1,k_2\ra$ for both the incoming photons from the left of the TLS, i.e., $A_2=V_1^2/V^2,B_2=V_1V_2/V^2$, and $C_2=V_2^2/V^2$. 
\bea
\la 1;k_1,k_2|I|1;k_1,k_2\ra&=&-\f{\mathcal{L}}{2\pi^2}\f{V_1^2V_2^2}{V^3}{\rm Im}[e_{k_1}+e_{k_2}]\nn \\&+&\f{V_1^4V_2^2}{\pi^2 V^5} {\rm Im}\Big[e_{k_2}^2e_{k_1}^*+e_{k_1}^2e_{k_2}^*\Big].
\label{twocurr1}
\eea
Here $\mathcal{L}$ is the total length of the full system. The first term in Eq.(\ref{twocurr1}) is the contribution from two noninteracting photons and is similar to Eq.(\ref{sincurr}). The last part in Eq.(\ref{twocurr1}) is the change in two-photon current (call $\delta I(1;k_1,k_2)$) due to photon-photon interactions and $\delta I(1;k_1,k_2)$ comes from the two-photon bound state part of the full wave-function. Also note that $\delta I(1;k_1,k_2)$ is a factor $\mathcal{L}$ smaller than the noninteracting current; this signifies that the probability of finding two photons near the TLS at any time is order of $1/\mathcal{L}$ for a 1D system of size $\mathcal{L}$ with two photons. $\mathcal{L}$ will disappear in the multi-photon-current in the system \cite{Dhar08}. We can have $\delta I(1;k_1,k_2)$ being similar in magnitude of noninteracting current for an optimal two-photon interaction. This occurs when two photons are incident upon the TLS within a time delay of absorption and spontaneous emission by the TLS. We plot $\delta I(1;k_1,k_2)$ vs $\Omega$ for a fixed energy of incident photons in Fig.\ref{pl2}. We find from Fig.\ref{pl2} that $\delta I(1;k_1,k_2)$ changes sign near the single-photon resonance energy, i.e., $\Omega=k_1$ or $k_2$. A positive $\delta I(1;k_1,k_2)$ can be interpreted as bunching of two photons after scattering where as a negative $\delta I(1;k_1,k_2)$ is antibunching.
\begin{figure}
\includegraphics[width=8cm]{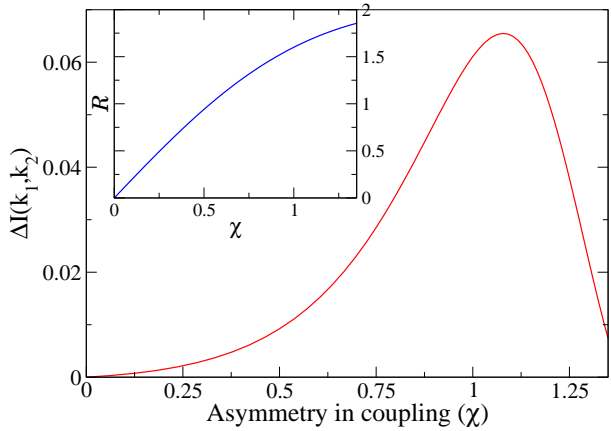}
\caption{(color online) Plot of $\Delta I(k_1,k_2)$ vs asymmetry in coupling $\chi$. Inset shows rectification $\mathcal{R}$ vs $\chi$. Here $k_1=0.3$, $k_2=0.7$, $V_2=0.15$, and $\Omega=0.5$.}
\label{pl3}
\end{figure}

We similarly derive the current for two photons  being incident from the right side of the TLS, i.e., $A_2=V_2^2/V^2,B_2=-V_1V_2/V^2,$ and $C_2=V_1^2/V^2$. 
\bea
\la 2;k_1,k_2|I|2;k_1,k_2\ra&=&\f{\mathcal{L}}{2\pi^2}\f{V_1^2V_2^2}{V^3}{\rm Im}[e_{k_1}+e_{k_2}]\nn \\&-&\f{V_1^2V_2^4}{\pi^2 V^5} {\rm Im}\Big[e_{k_2}^2e_{k_1}^*+e_{k_1}^2e_{k_2}^*\Big].
\label{twocurr2}
\eea
 We see that the contribution from two noninteracting photons (the first part in Eq.(\ref{twocurr2})) is similar in magnitude to that in Eq.(\ref{twocurr1}). But the change in two-photon current $\delta I(2;k_1,k_2)$ (last part in Eq.(\ref{twocurr2})) is different in magnitude from $\delta I(1;k_1,k_2)$ for $V_1 \ne V_2$. We plot $-\delta I(2;k_1,k_2)$ versus $\Omega$ in Fig.\ref{pl2}. Hence we find that $|\la 1;k_1,k_2|I|1;k_1,k_2\ra| \ne |\la 2;k_1,k_2|I|2;k_1,k_2\ra|$ for $V_1 \ne V_2$, i.e., the directly coupled 1D waveguide-TLS system acts as an optical diode for two-photon transport.  The mechanism behind diode like behavior in the present system is the redistribution of energy and momentum of photons after passing through the TLS. The asymmetry in the coupling generates asymmetry in the distribution of energy and momentum of photons  coming from the left or right of the TLS. In fact, $|\delta I(1;k_1,k_2)|:|\delta I(2;k_1,k_2)|=V_1^2/V_2^2$ is exactly same to the ratio of the bound state contributions for both the incoming photons from the left or right of the TLS. We remind that the redistribution of energy and momentum of scattered photons occurs via the two-photon bound states. 
A similar phenomenon for electrons has been studied recently \cite{Roy09}. Note that the present system acts as a diode in the fully quantum regime for a minimal two photons. It also differs from the classical spin-boson thermal rectifier \cite{Segal05} operating at a finite temperature bias. We here define $\Delta I(k_1,k_2)=|\la 1;k_1,k_2|I|1;k_1,k_2\ra|-|\la 2;k_1,k_2|I|2;k_1,k_2\ra|$, and the dimensionless rectification, $\mathcal{R}=2 \Delta I(k_1,k_2)/(|\delta I(1;k_1,k_2)|+|\delta I(2;k_1,k_2)|)$. We also quantify the asymmetry in the coupling by $\chi=2(V_1-V_2)/(V_1+V_2)$. We plot $\Delta I(k_1,k_2)$ and $\mathcal{R}$ with $\chi$ for a fixed incident energy of two photons and a fixed $\Omega$ in Fig.\ref{pl3}. We use a constant $V_2$ and change $V_1$ in Fig.\ref{pl3}. We find that $\Delta I(k_1,k_2)$ first increases with $\chi$ and then falls after a maximum value. On the contrary, $\mathcal{R}$  increases monotonically with $\chi$ and finally saturates at its maximum value 2. The multi-photon transport is also correlated in this system and it acts as an optical diode for many photons too. The technique employed here can be used to find current with three or more incoming photons. 

The proposed optical diode may have potential applications to build quantum circuits for optical quantum information processing and quantum computation \cite{Bouwmeester00}. However, the practical realization of the optical diode depends on achieving strong and asymmetric coupling between the TLS and the photons in the waveguide. Strong light-matter interaction is the main focus of much research in quantum optics for quite sometime and has been demonstrated successfully in the recent past \cite{Wallraff04, Hennessy07, Akimov07}. The asymmetry in coupling has not been much discussed before. The optical diode can be fabricated either with a TLS in a single waveguide or with a TLS being coupled to two different waveguides. It is technically challenging to fix a TLS at the right location in a single waveguide \cite{Wallraff04, Hennessy07, Fushman08}. Again there may be high insertion loss for a TLS being coupled to two different waveguides. Recently an ultrahigh-Q coupled resonator optical waveguides with low losses has been realized \cite{Notomi08}. A strong direct coupling between an emitter and surface plasmons in a nanowire has also been achieved \cite{Akimov07}. Hence it seems to be possible to reduce the insertion loss. In fact, the emitter-surface plasmons system is a promising candidate for realizing asymmetry in coupling as the coupling between the emitter and surface plasmons is completely geometrical and depends on the diameter of the conducting nanowire \cite{Chang2}.    

The discussions with S. Mookherjea, D. I. Schuster and A. Wallraff on the experimental realization of asymmetric coupling are gratefully acknowledged. The work has been funded by the DOE under Grant No. DE-FG02-05ER46204.

\end{document}